\documentclass[manuscript]{aastex62}

\newcommand{\kepler}{{\it Kepler}}
\newcommand{\tess}{{\it TESS}}

\submitjournal{ApJ, 2019 Jan 31}
\revised{2019 Apr 2}
\accepted{2019 Apr 20}

\shorttitle{Variability of Kepler-76b}
\shortauthors{Jackson, Adams, Sandidge, Kreyche, \& Briggs}

\usepackage{hyperref}
\usepackage{graphics,graphicx}

\begin{document}

\title{Variability in the Atmosphere of the Hot Jupiter Kepler-76b}

\author{Brian Jackson}
\affiliation{Department of Physics, Boise State University, 1910 University Drive, Boise ID 83725-1570 USA}

\author{Elisabeth Adams}
\affiliation{Planetary Science Institute, 1700 E. Ft. Lowell, Suite 106, Tucson, AZ 85719, USA}

\author{Wesley Sandidge}
\affiliation{Department of Physics, Boise State University, 1910 University Drive, Boise ID 83725-1570 USA}

\author{Steven Kreyche}
\affiliation{Department of Physics, Boise State University, 1910 University Drive, Boise ID 83725-1570 USA}
\affiliation{Department of Physics, University of Idaho, Moscow, Idaho 83844-0903, USA}

\author{Jennifer Briggs}
\affiliation{Department of Physics, Boise State University, 1910 University Drive, Boise ID 83725-1570 USA}

\begin{abstract}
Phase curves and secondary eclipses of gaseous exoplanets are diagnostic of atmospheric composition and meteorology, and the long observational baseline and high photometric precision from the \kepler\ Mission make its dataset well-suited for exploring phase curve variability, which provides additional insights into atmospheric dynamics. Observations of the hot Jupiter Kepler-76b span more than 1,000 days, providing an ideal dataset to search for atmospheric variability. In this study, we find that Kepler-76b's secondary eclipse, with a depth of $87 \pm 6$ parts-per-million (ppm), corresponds to an effective temperature of 2,830$^{+50}_{-30}$ K. Our results also show clear indications of variability in Kepler-76b's atmospheric emission and reflectivity, with the phase curve amplitude typically $50.5 \pm 1.3$ ppm but varying between 35 and 70 ppm over tens of days. As is common for hot Jupiters, Kepler-76b's phase curve shows a discernible offset of $\left( 9 \pm 1.3 \right)^\circ$ eastward of the sub-stellar point and varying in concert with the amplitude. These variations may arise from the advance and retreat of thermal structures and cloud formations in Kepler-76b's atmosphere; the resulting thermal perturbations may couple with the super-rotation expected to transport aerosols, giving rise to a feedback loop. Looking forward, the \tess\ Mission can provide new insight into planetary atmospheres, with good prospects to observe both secondary eclipses and phase curves among targets from the mission. \tess's increased sensitivity in red wavelengths as compared to \kepler\ means that it will probably probe different aspects of planetary atmospheres.

\end{abstract}

\keywords{planets and satellites: atmospheres -- planets and satellites: gaseous planets -- planets and satellites: individual (Kepler-76b)}

\section{Introduction}
\label{sec:Introduction}
The age of exoplanet meteorology has arrived -- from the first detection of atmospheric thermal emission \citep{2005Natur.434..740D, 2005ApJ...626..523C} to the recent detection of the secondary eclipses of an Earth-sized planet \citep{2016Natur.532..207D}, observation, characterization, and modeling of exoplanet atmospheres has blossomed into a rich field approaching that of solar system meteorology in scope and complexity. Indeed, the much wider range of dynamical, compositional, and radiative environments across which exoplanets have been discovered means the study of these systems will undoubtedly provide insight into the planetary atmospheres in our own solar system.

Among the first key results in the study of exoplanet meteorology was the mapping of atmospheric emission from the transiting hot Jupiter HD189733b \citep{2007Natur.447..183K}. In that study, observations in the \emph{Spitzer} Space Telescope's 8 $\mu$m band revealed a deep secondary eclipse (0.3381\% in differential flux) corresponding to a dayside brightness temperature of more than 1,200 K. Perhaps of more interest to meteorological studies, the data also showed signs of a planetary emission during the entire orbit, peaking as the tidally locked planet's dayside rotated into view. However, instead of being centered around the eclipse, the planet's phase curve peak occurred about 2 hours before, indicating the brightest and hottest region in the atmosphere was shifted about 16$^\circ$ eastward from the sub-stellar point. \citet{2007Natur.447..183K} attributed this offset to advective transport within the atmosphere, a result anticipated by earlier studies of hot Jupiter atmospheres \citep[e.g.,][]{2002A&A...385..166S}. 

The mountain of data made available by the \kepler\ Mission has allowed analysis of phase curves for numerous planets \citep[e.g.,][]{2015ApJ...804..150E}. However, since the \kepler\ data involve visible wavelengths (with a bandpass spanning 400 to 900 nm), these phase curves likely involve a combination of thermal emission and reflected light from high-altitude aerosols \citep{2000ApJ...538..885S}. The aerosols obscure the lower atmosphere, complicating the determination of atmospheric composition \citep{2016Natur.529...59S}. 

Although the composition, size distribution, and vertical mixing of aerosols are difficult to determine \emph{a priori}, their influence on the planetary phase curve, in particular the resulting phase curve offset, may be diagnostic. \citet{2016ApJ...828...22P} summarize the distribution of observed phase curve offsets and highlight a clear correlation with planetary effective temperature. Cooler planets show offsets toward the west, consistent with the expectation that clouds dominate these planets' phase curves. By contrast, some hotter planets, such as HD189733 b, show eastward offsets, symptomatic of either homogeneous or little dayside cloud cover and downwind transport of thermally emitting regions, although there are important exceptions, such as CoRoT-2b \citep{2018NatAs...2..220D}. \citet{2016ApJ...828...22P} also suggest that the same super-rotating winds that drive hot regions eastward can draw aerosols that form on the nightside across a hot planet's western (night-to-day) terminator, forming a dayside cloud front whose eastern-most boundary depends on the aerosol composition -- the more refractory the aerosols, the farther east the boundary. These clouds can contribute to the planetary phase curve and, since they may be confined to the western hemisphere, they may compete with eastern-hemisphere emission and drive the observed phase curve offset toward the west.

All these effects take place in a highly dynamic atmosphere, in which winds approach sonic speeds \citep{2018ApJ...853..133K}, and the interplay between transport of aerosols, which act to cool the atmosphere, and potent stellar insolation, which drives the winds, produce variable atmospheric structures. This variability likely shows up in the reflected and emitted light from the planets' atmospheres. Studying aerosol transport in the atmosphere of HD209458b, \citet{2013A&A...558A..91P} predicted variations in the secondary eclipse depths as large as 50\% in IR wavelengths over tens of days. Analyzing hundreds of days of \kepler\ observations of the hot Jupiter HAT-P-7 b, \citet{2016NatAs...1E...4A}, indeeed, found significant variations in the phase curve offset, which wandered back and forth across the sub-stellar point over tens of days, likely driven, at least in part, by variations in aerosol transport. Although \citet{2016NatAs...1E...4A} were not able to robustly detect them, variations in the phase curve amplitude probably accompanied the offset fluctuations since variations in aerosol transport would also be expected to modify the planet's total geometric albedo. \citet{2019ApJ...872L..27H} explored the effects of magnetic fields on hot Jupiter phase curves and found that planetary-scale equatorial magneto-Kelvin waves could drive westward tilting eddies and produce the observed periodic westward offset in HAT-P-7b's phase curve reported in \citet{2016NatAs...1E...4A}.

In principle, it should be possible to connect variations in a planet's atmospheric reflection and emission to phase curve variability, but this requires a large enough population of planets that spans a range of conditions to piece together a cogent story. Following this thread, we analyzed \kepler\ observations of the hot Jupiter Kepler-76b, a two Jupiter-mass (${\rm M_{Jup}}$) planet with an orbital period of 1.5 days around a hot (6,300 K) star \citep{2013ApJ...771...26F}. In addition to having a clear atmospheric signal, Kepler-76b also induces observable ellipsoidal variations on its host star (the photometric signature of tidal bulges on the star rotating in and out of view -- \citealp{2010ApJ...713L.145W}) and a beaming signal due to the radial velocity of the star \citep{2003ApJ...588L.117L}. With nearly 1,400 days of observation, we find significant variability in the planet's phase curve amplitude, $A_{\rm planet}$, and offset, $\delta$, qualitatively consistent with the variations seen for HAT-P-7 b. We have also updated the transit parameters for Kepler-76b.

This article is organized as follows: in Section \ref{sec:Data_Analysis}, we describe our process for conditioning the \kepler\ dataset and then our methodology for modeling Kepler-76b's transit, eclipse, and other photometric signals. In Section \ref{sec:Results}, we discuss our results, and in Section \ref{sec:Discussion}, we discuss their implications and future prospects for using \tess\ data to conduct similar analyses.

\section{Data Analysis}
\label{sec:Data_Analysis}
\subsection{Data Conditioning}
\begin{figure}
\includegraphics[width=\textwidth]{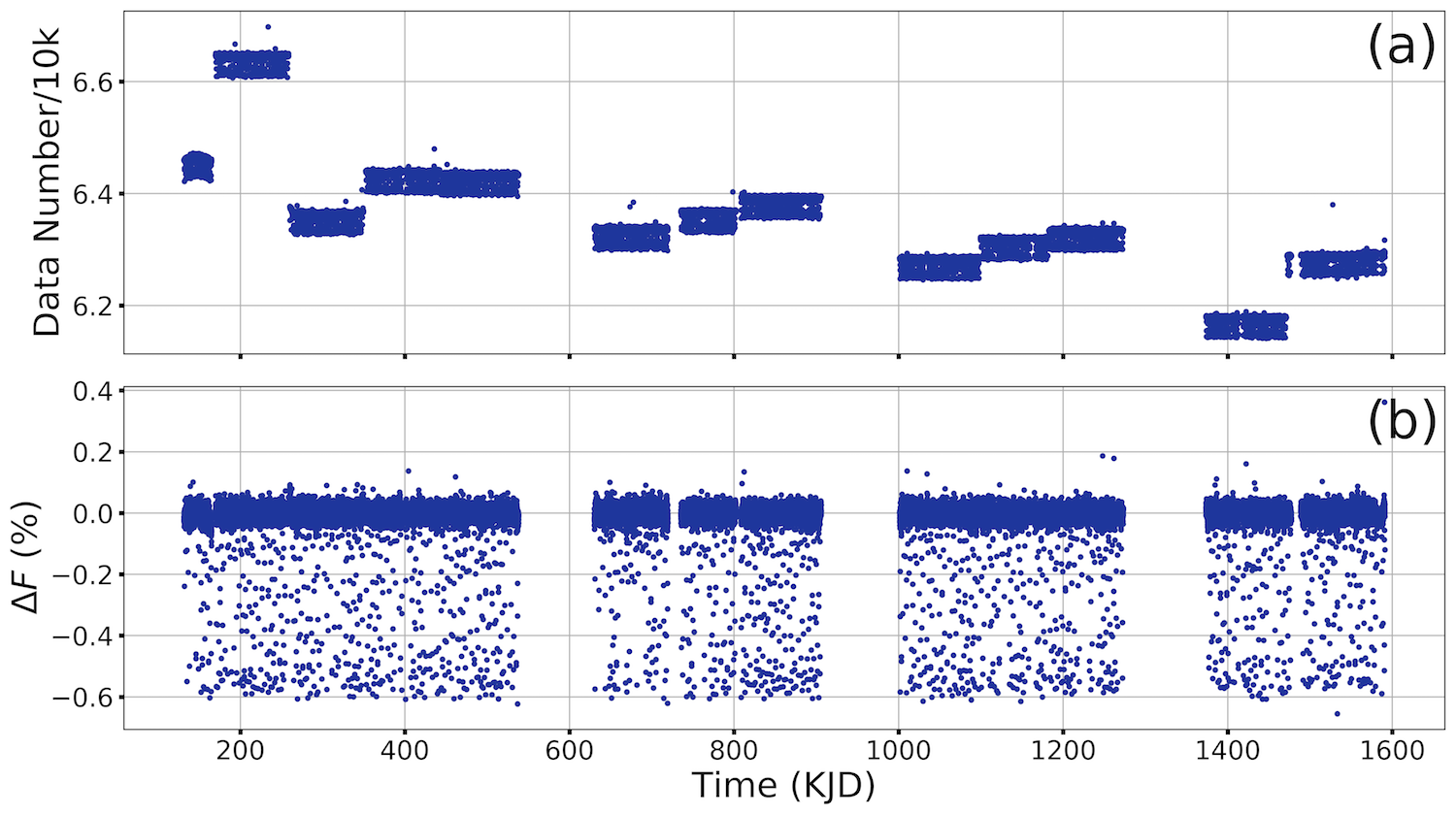}
\caption{The quarter-by-quarter (a) raw and (b) conditioned PDCSAP\_FLUX. Time along the x-axis is shown in the \kepler\ mission's barycentric Julian date minus 2454833 (midnight on 2009 January 1), and the y-axes show variations in flux.\label{fig:raw-conditioned-data_Analysis_of_Kepler76b}}
\end{figure}

We applied the following steps to condition all fourteen available quarters (Quarters 1-5, 7-9, 11-13, 15-17) of \kepler\ long-cadence (i.e., 30-min integration times) data:
\begin{enumerate}
\item To reduce variations in each quarter's data to the same scale, we subtracted the median value from each quarter's PDCSAP\_FLUX time-series before dividing through by that same median value.
\item We next applied a median boxcar filter with a window size equal to four orbital periods. We experimented with window sizes from 1-15 orbital periods (holding fixed the transit parameters to those reported in \citealt{2013ApJ...771...26F}), then fitting the out-of-transit photometric signals described below using the Levenberg-Marquardt algorithm (LM) \citep{newville_2014_11813}, including the planet's eclipse and phase curve; four orbital periods maximized the eclipse depth and minimized the distortion to the phase curve using all available data. To mitigate edge-effect distortions from our boxcar filter, we extended the time-series out a full window length beyond both ends by reflecting the original time-series across its boundary.
\item Finally, we stitched each quarter's conditioned data into one long time-series.
\end{enumerate}
Figure \ref{fig:raw-conditioned-data_Analysis_of_Kepler76b} illustrates a portion of the raw and conditioned time-series for Kepler-76b. 

\subsection{Analyzing All Data Folded Together}
\begin{deluxetable}{lcr}
\tablecaption{Transit, Eclipse, and Other Photometric Variation Model Parameters\label{tbl:Model_Parameters}}
\tabletypesize{\footnotesize}
\startdata
Transit & & \\
\hline
$T_0$ (BJD) &   $2454965.00396 \pm 5 \times 10^{-5}$ & Time of Primary Transit\\
Period (days) & $1.54492871 \pm 9 \times 10^{-8} $ & Orbital Period\\
$R_{\rm p}/R{\star}$ & $0.085 \pm 0.001$ & Planetary Radius/Stellar Radius\\
$a/R_{\star}$ & $5.103^{+0.058}_{-0.060}$ & Orbital Semi-Major Axis/Stellar Radius\\
$b$ & $0.908 \pm 0.003$ & Impact Parameter\\
$i$ (degrees) & $79.7 \pm 0.1$ & Orbital Inclination \\
$(\gamma_1, \gamma_2)$ & (0.313, 0.304) & Limb-Darkening Coefficients \citep{2013ApJ...771...26F}\\
\hline
\hline
Eclipse/Photometric Variations & & \\
\hline
$D$ (ppm) & $87 \pm 6$ & Secondary Eclipse Depth\\
$A_{\rm planet}$ (ppm) & $50.5 \pm 1.3$ & Phase Curve Amplitude\\
$\delta$ (degrees) & $-9.0 \pm 1.3$ & Phase Curve Offset\\
$A_{\rm ellip}$ (ppm) & $13 \pm 1$ & Ellipsoidal Variation (EV) Amplitude\\
$A_{\rm beam}$ (ppm) & $3.8 \pm 0.3$ & Beaming Effect Amplitude\\
\hline
\hline
Other Parameters & & \\
\hline
$M_{\star}\ ({\rm M_{\odot}})$ & $1.2 \pm 0.2$ & Stellar Mass \citep{2013ApJ...771...26F}\\
$g$ & 0.07 & Gravity-Darkening Parameter \citep{2011AA...529A..75C} \\
$\alpha_{\rm beam}$ & 0.92 & Beaming Coefficient \citep{2013ApJ...771...26F}\\
$\alpha_{\rm ellip}$ & 1.02 & EV Coefficient \\
$K_{\rm z}$ (km/s) & $0.306 \pm 0.020$ & Radial Velocity \citep{2013ApJ...771...26F}\\
$q_{\rm ellip}$ & $\left( 1.9 \pm 0.2 \right) \times 10^{-3}$ & Planet-Star EV Mass Ratio \\
$M_{\rm p, ellip}\ ({\rm M_{\rm Jup}})$ & $2.4 \pm 0.3$ & Planetary EV Mass \\
$q_{\rm beam}$ & $\left( 1.7\pm 0.1 \right) \times 10^{-3}$ & Planet-Star Beaming Mass Ratio \\
$M_{\rm p, beam}\ ({\rm M_{\rm Jup}})$ & $2.1 \pm 0.2$ & Planetary Beaming Mass
\enddata
\end{deluxetable}

After applying the above conditioning procedure, we fit a series of photometric models to all the available data. We first analyzed the out-of-transit portion of Kepler-76b's light curve since those signals (the ellipsoidal variations, the planet's phase curve, and the Doppler beaming signal) are independent of the transit and can, in principle, contribute to the transit portion (although we found their effects are negligible). 

We folded all data on the best-fit period reported by \citet{2013ApJ...771...26F} and masked out the transit and eclipse portions of the light curve and fit the following model:
\begin{equation}
    % \Delta F & = & F_0 - A_{\rm ellip} \cos \left(2\times2\pi\phi \right) + A_{\rm beam} \sin \left(2 \pi\phi \right) - A_{\rm planet} \cos \big(2\pi \left( \phi - \delta \right)\big)\nonumber \\
    \Delta F = F_0 - A_{\rm ellip} \cos \left(2\times2\pi\phi \right) + A_{\rm sin} \sin \left(2 \pi\phi \right) - A_{\rm cos} \cos \left(2\pi \phi \right),
\label{eqn:BEER_curve}
\end{equation}
where $F_0$ represents a constant baseline and $\phi$ the orbital phase ($\phi = 0$ at mid-transit). The second term represents the ellipsoidal variations induced by the planet's tidal gravity \citep{2010ApJ...713L.145W}, with $A_{\rm ellip}$ its amplitude. The third and fourth terms are a combination of the Doppler beaming signal \citep{2003ApJ...588L.117L} and the planet's reflected/thermally emitted phase curve, with allowance for an offset in the curve's maximum from superior conjunction at $\phi = 0.5$ \citep{2013ApJ...771...26F}. We also allowed the per-point uncertainty (noise) to be a free parameter and estimated the model likelihood $L$ as
\begin{equation}
    \ln L = -\frac{1}{2} \sum_i \bigg( \frac{ d(t_{\rm i}) - m(t_{\rm i}) }{\sigma} \bigg)^2 + \ln \left( 2 \pi \sigma^2 \right),
    \label{eqn:likelihood}
\end{equation}
where $d(t_{\rm i})$ is the datum at time $t_{\rm i}$, $m(t_{\rm i})$ the model for that same time, and $\sigma$ the per-point uncertainty. With this likelihood, we used a Markov Chain Monte Carlo (MCMC) analysis \citep{2013PASP..125..306F} with 100 walker chains, each with 5,000 links and a burn-in phase of 2,500 links. We aimed for convergence by requiring small auto-correlation times \citep[e.g.,][]{geyer1992} for the mean chain of each fit parameter (none exceeded 80 links) and all Geweke Z-scores \citep{Geweke92evaluatingthe} of about 3 or less. 

Traditionally, the beaming and planetary phase curve signals are treated separately through the combination $A_{\rm beam} \sin \left(2 \pi\phi \right) - A_{\rm planet} \cos \big(2\pi \left( \phi - \delta \right)\big)$. This expression can be re-cast using $A_{\rm sin} = A_{\rm beam} - A_{\rm planet} \sin \left( 2 \pi \delta \right)$ and $A_{\rm cos} = A_{\rm planet} \cos \left( 2 \pi \delta \right)$. Since the beaming and planetary phase curve signals have the same frequency (once per orbit) and the phase offset $\delta$ allows them to phase up (at least in principle), they are degenerate; completely independent fits for $A_{\rm beam}$, $A_{\rm planet}$, and $\delta$ are not possible.

\begin{figure}
\includegraphics[width=\textwidth]{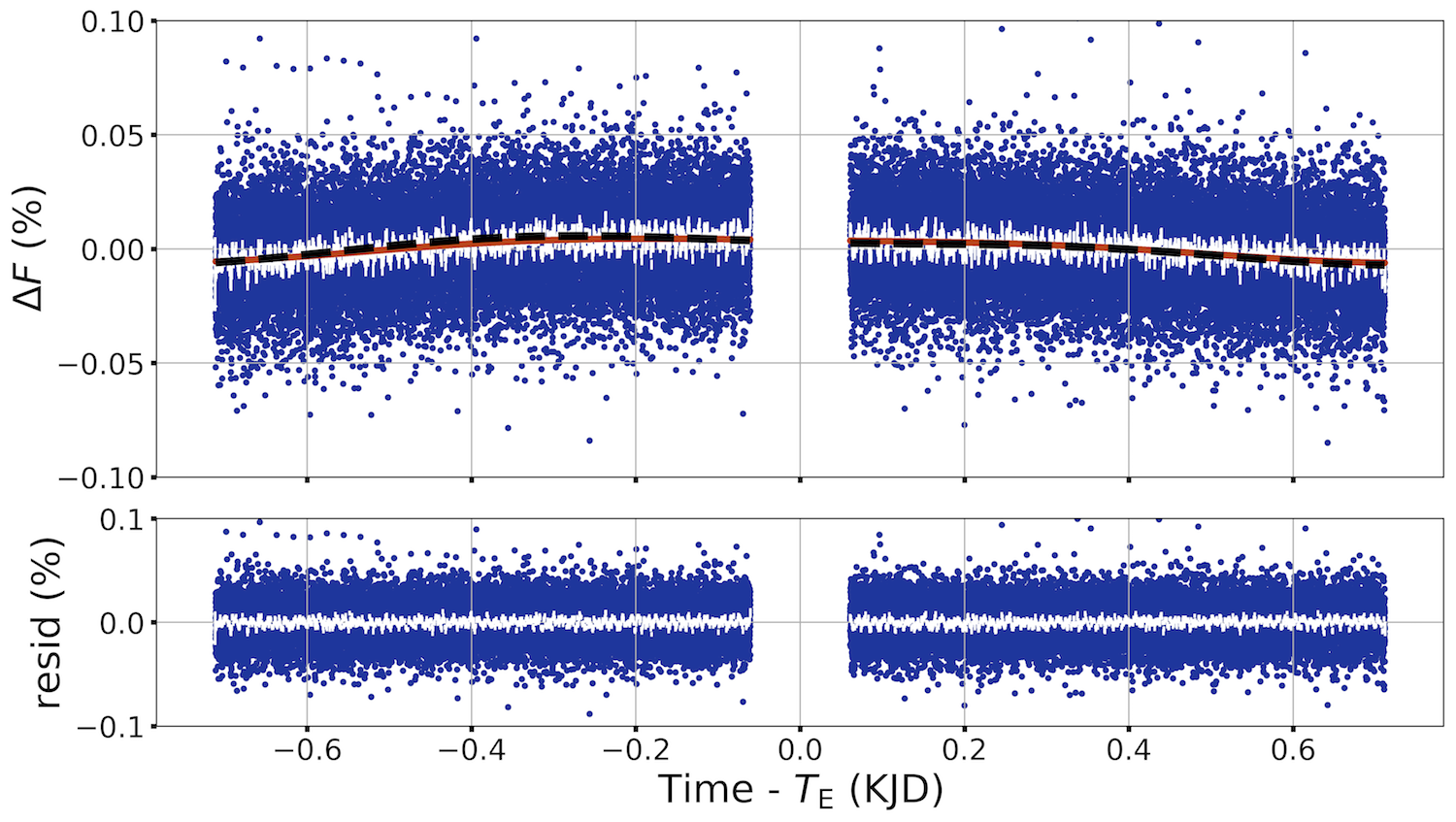}
\caption{The blue points in the top panel show photometric measurements for the Kepler-76 system, outside of the planet's eclipse and transit, folded on the ephemeris reported as in \citet{2013ApJ...771...26F}. The white points show these same data binned in 1-min-wide bins, with error bars showing the standard deviation in each bin. The solid orange line shows our best-fit model (Equation \ref{eqn:BEER_curve}), while the dashed black line shows that from \citet{2013ApJ...771...26F}. The bottom panel shows the residuals between our model and the data.\label{fig:BEER-curve-fit_Analysis_of_Kepler76b}}
\end{figure}

\begin{figure}
\includegraphics[width=\textwidth]{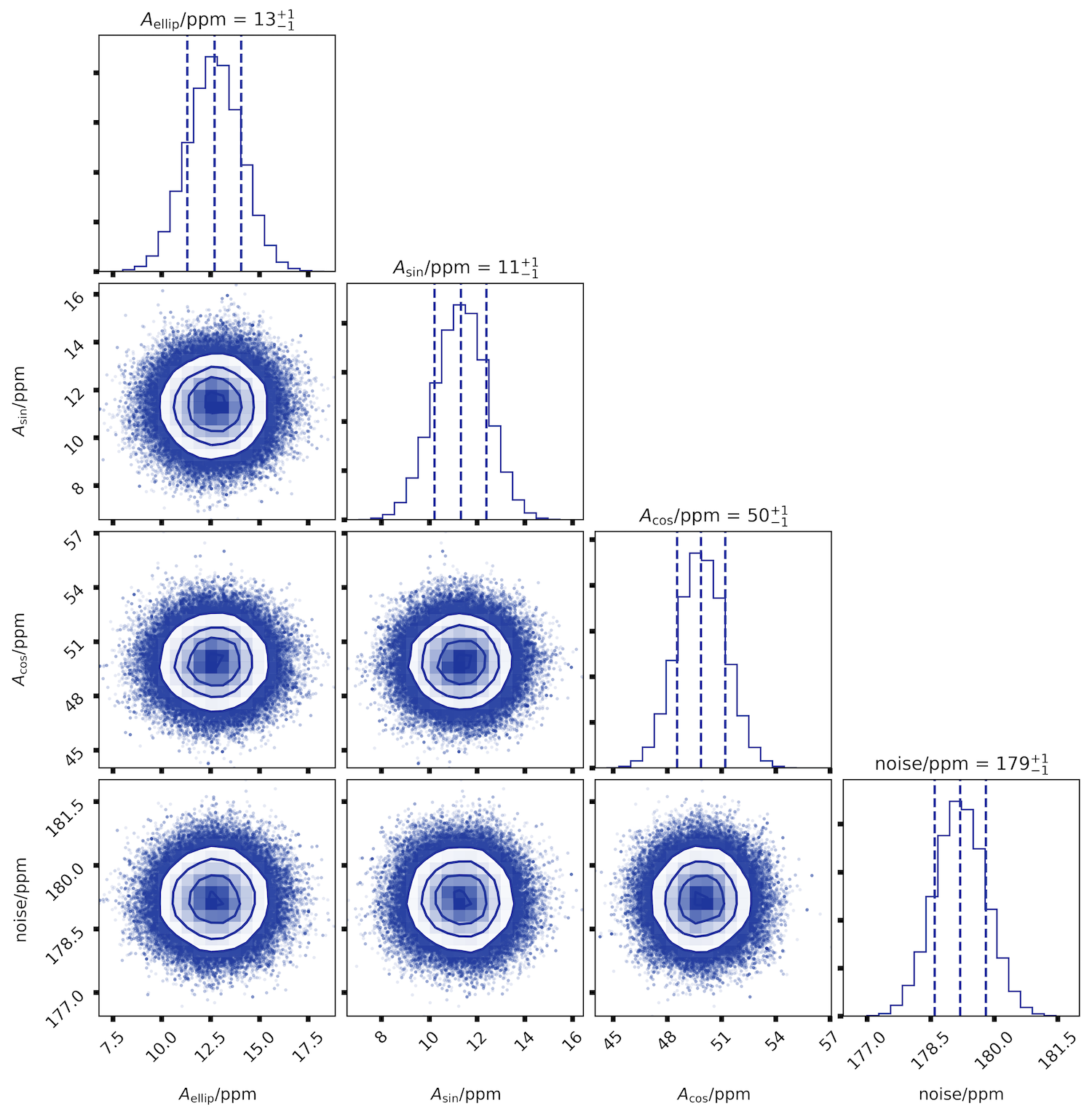}
\caption{The posterior distributions for the best-fit parameters for the model described by Equation \ref{eqn:BEER_curve} and corresponding to the orange curve in Figure \ref{fig:BEER-curve-fit_Analysis_of_Kepler76b}. The histograms along the top-right of each row/column shows the distribution for a parameter marginalized over all the other parameters, while distributions in the shaded contour plots are marginalized over the parameters not labeled to illustrate the correlations between parameters, which appear to be negligible. \label{fig:BEER-curve-fit-params_Analysis-of-Kepler76b}}
\end{figure}

\begin{figure}
\includegraphics[width=\textwidth]{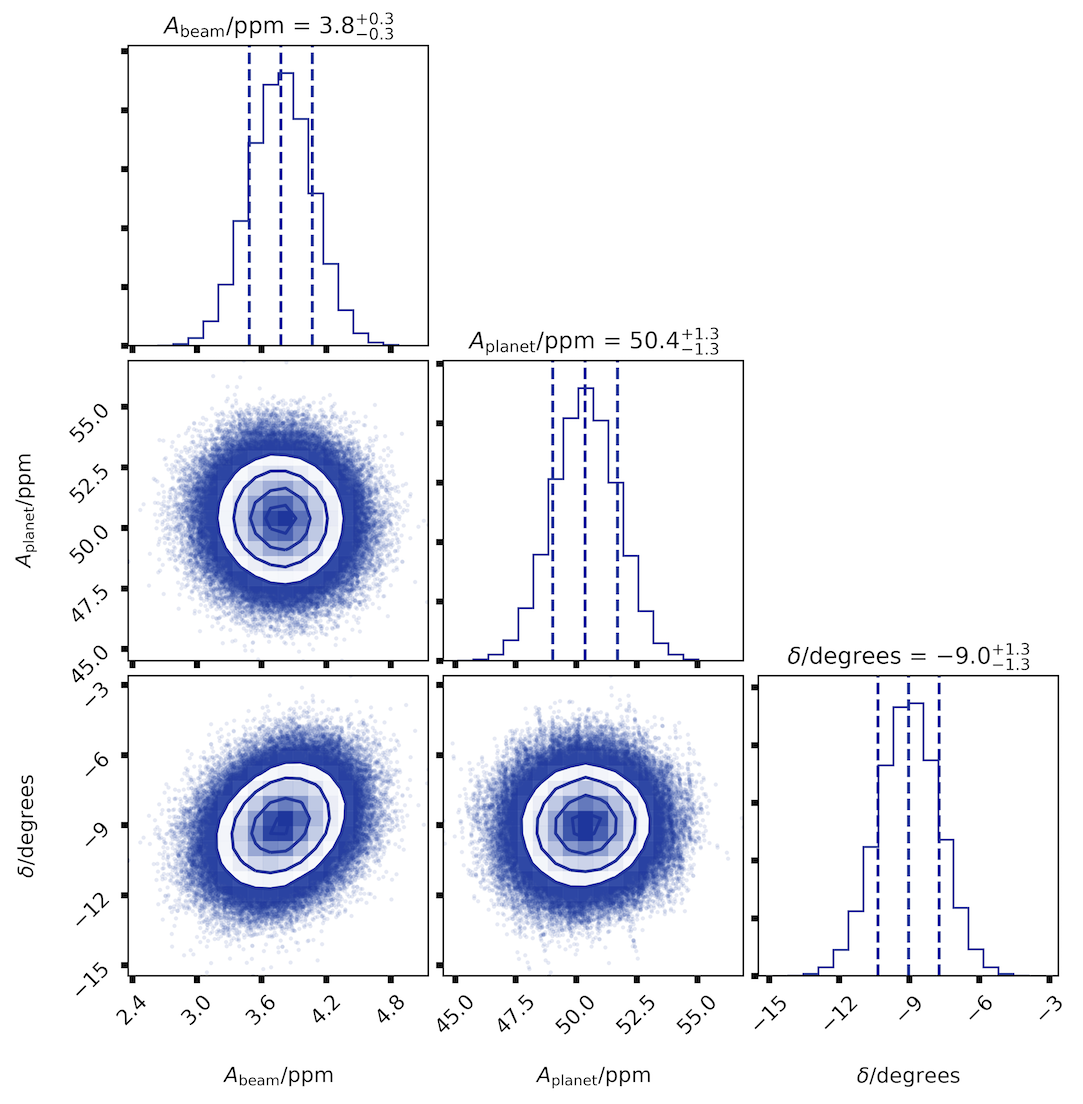}
\caption{The posterior distributions for the amplitude of the planet's phase curve and its phase offset, assuming the beaming signal amplitude implied by the radial velocity semi-amplitude reported in \citet{2013ApJ...771...26F}. \label{fig:Aplanet-delta-fit-params_Analysis-of-Kepler76b}}
\end{figure}

Figure \ref{fig:BEER-curve-fit_Analysis_of_Kepler76b} shows the fit and residuals for Equation \ref{eqn:BEER_curve} to the dataset. Figure \ref{fig:BEER-curve-fit-params_Analysis-of-Kepler76b} shows the resulting best-fit parameters, while Figure \ref{fig:Aplanet-delta-fit-params_Analysis-of-Kepler76b} shows the corresponding range of values for $A_{\rm planet}$ and $\delta$. To generate the latter figure, we calculated $A_{\rm beam}$ from the radial-velocity semi-amplitude $K_{\rm z} = 0.306 \pm 0.020$ km/s reported in \citet{2013ApJ...771...26F} using $A_{\rm beam} = 4\ \alpha_{\rm beam}$ and $K_{\rm z}/c = 3.8 \pm 0.3$ ppm, where $c$ is the speed of light \citep{2003ApJ...588L.117L} -- we did not use the $A_{\rm beam}$-value reported in \citet{2013ApJ...771...26F}. Then, we Monte-Carlo-sampled 250,000 times from a Gaussian distribution for $A_{\rm beam}$ (with the mean and width given above), as well as the MCMC chains for $A_{\rm sin}$ and $A_{\rm cos}$. Figure \ref{fig:Aplanet-delta-fit-params_Analysis-of-Kepler76b} shows the clear correlation between $A_{\rm beam}$ and $\delta$. As a final check, we fit the out-of-transit/out-of-eclipse data with an LM approach, holding $A_{\rm beam}$ fixed at 13.5 ppm, as reported in \citet{2013ApJ...771...26F}, and were able to recover the reported $A_{\rm planet}$ but not the $\delta$-value. 

Next, we analyzed the in-transit portion of the light curve. Importantly, we accounted for \kepler's finite integration time of 30-min-per-point by super-sampling each point, i.e. we modeled the light curve at a cadence of 3-min-per-point and then downsampled to 30-min-per-point \citep[cf.][]{2010MNRAS.408.1758K} (the finite integration had no significant effect on the Equation \ref{eqn:BEER_curve} model). By first subtracting the Equation \ref{eqn:BEER_curve} model from the in-transit portion of the light curve, we removed its (very small) influence on the transit. Next, we used LM to fit a standard transit curve with quadratic limb-darkening \citep{2002ApJ...580L.171M}\footnote{As implemented in PyAstronomy -- \url{https://github.com/sczesla/PyAstronomy}.} to that portion of the time-series within two transit durations of the reported mid-transit time, allowing the out-of-transit baseline to vary. This analysis netted initial best-fit transit parameters: $a/R_\star$, the ratio of the planet's semi-major axis to the stellar radius; $b$, the impact parameter; and $R_{\rm p}/R_\star$, the planet-to-star radius ratio. We assumed the orbital eccentricity is equal to zero based on the photometric and radial velocity analyses in \citet{2013ApJ...771...26F} and held fixed the limb-darkening coefficients $\gamma_1 = 0.313$ and $\gamma_2 = 0.304$. We explored two other sets of quadratic limb-darkening coefficients, estimated assuming the stellar parameters from \citet{2013ApJ...771...26F} and using the model from \citet{2015MNRAS.450.1879E} but found different values had negligible impact on our transit analysis. 

\begin{figure}
    \includegraphics[width=\textwidth]{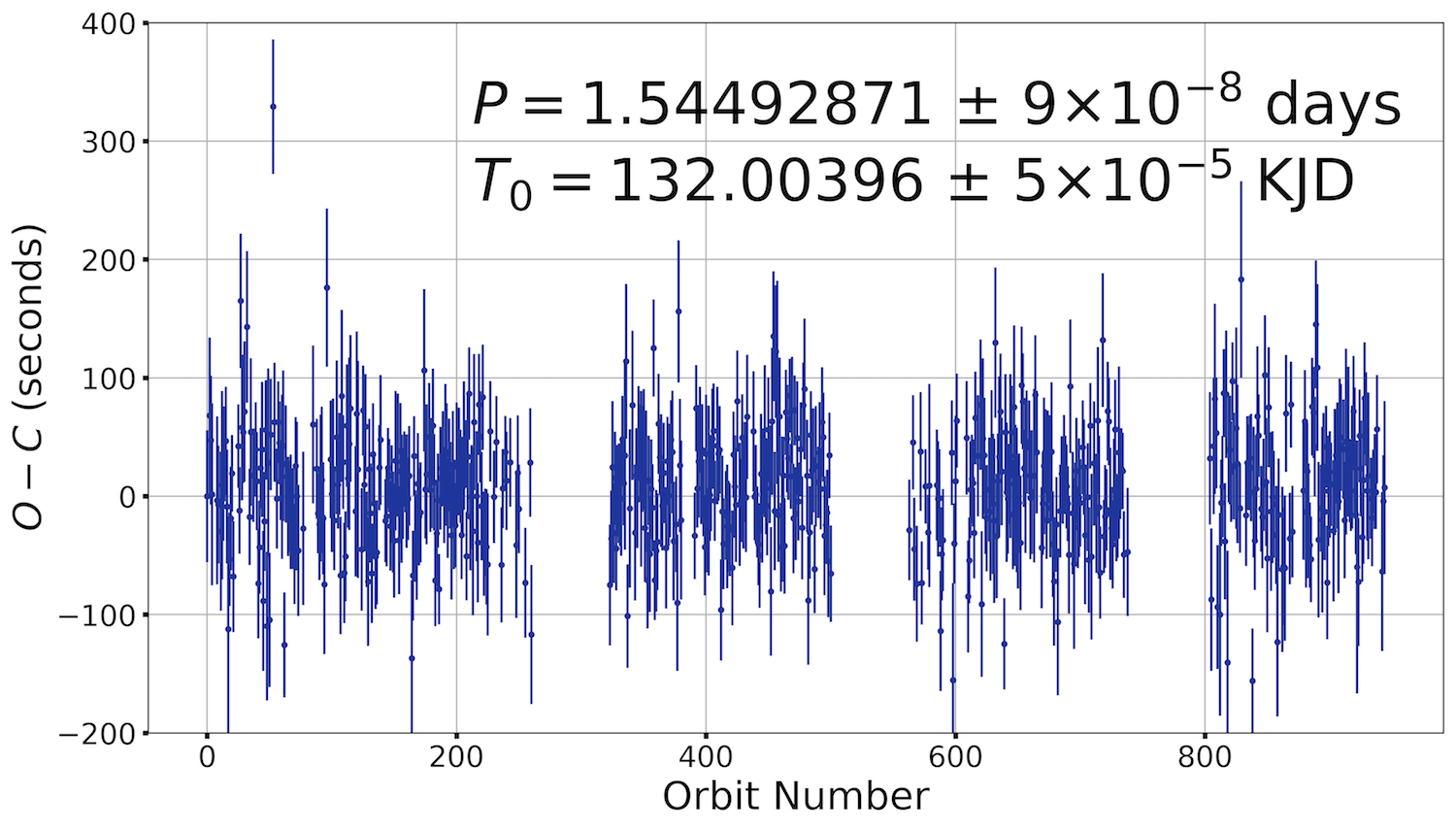}
    \caption{The observed mid-transit times ($O$) compared to those calculated ($C$) from a linear ephmeris. The orbital period $P$ and initial mid-transit time $T_0$ are shown. The gaps in the data near orbit numbers 300, 550, and 800 represent gaps in the available \kepler\ data.\label{fig:TTVs_Analysis_of_Kepler76b}}
\end{figure}

To check for transit-timing variations, we fit each individual transit (with the full orbital light curve this time and not just the in-transit portion) using LM again and held fixed all transit parameters except the mid-transit times and out-of-transit baseline. For uncertainties, we took the square root of the diagonal elements of the resulting covariance matrix \citep[p.~790]{Press:2007:NRE:1403886} and confirmed the accuracy of the mid-transit times and uncertainties by fitting transit curves with an MCMC analysis \citep{2013PASP..125..306F} to several individual transits -- we achieved convergence with 100 walkers, each a chain of 1,000 links and a burn-in phase of 300 links. Although most single transits are detectable, they usually comprise one or two data points and so provide no useful constraints on parameters other than the mid-transit times. Finally, we fit the collection of mid-transit times with both linear and quadratic ephmerides using an MCMC analysis (100 walker chains, 5,000 links in each with a burn-in of 1,500 links) and found no significant departure from the latter. We also found no significant periodicities using a Lomb-Scargle periodogram \citep{1976Ap&SS..39..447L, 1982ApJ...263..835S}. Figure \ref{fig:TTVs_Analysis_of_Kepler76b} shows the resulting transit ephemeris, which is consistent to within uncertainties with that reported in \citet{2013ApJ...771...26F}. 

\begin{figure}
    \includegraphics[width=\textwidth]{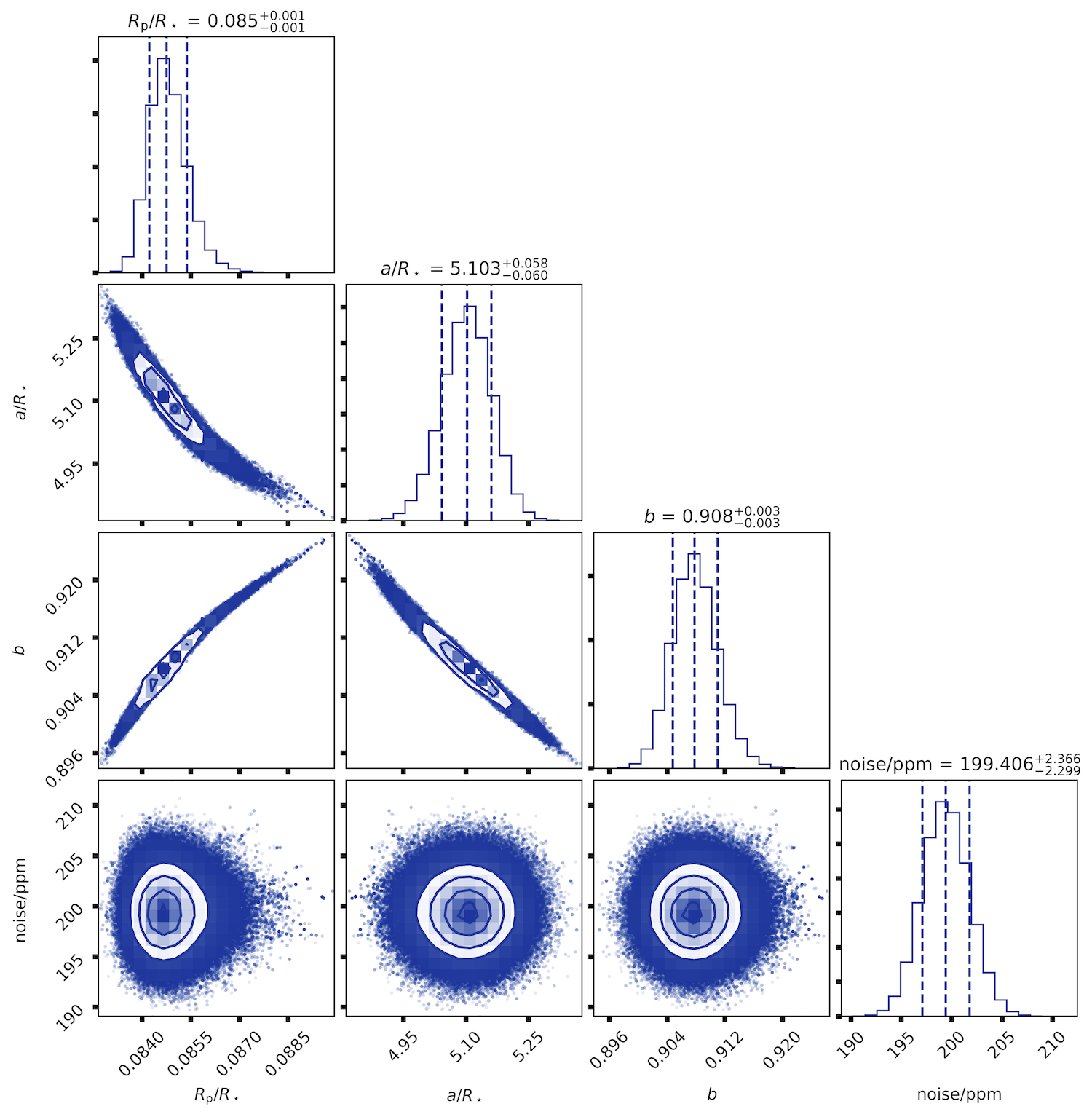}
    \caption{The distributions of best-fit parameters for all the data phase-folded on the ephemeris shown in Figure \ref{fig:TTVs_Analysis_of_Kepler76b}: the planet-star radius ratio $R_{\rm p}/R_\star$, the ratio of the planet's orbital semi-major axis to the stellar radius $a/R_\star$, the impact parameter $b$, and an estimate of the per-point noise in parts-per-million.}
    \label{fig:folded-transit-corner-plot_Analysis-of-Kepler76b}
\end{figure}

\begin{figure}
    \includegraphics[width=\textwidth]{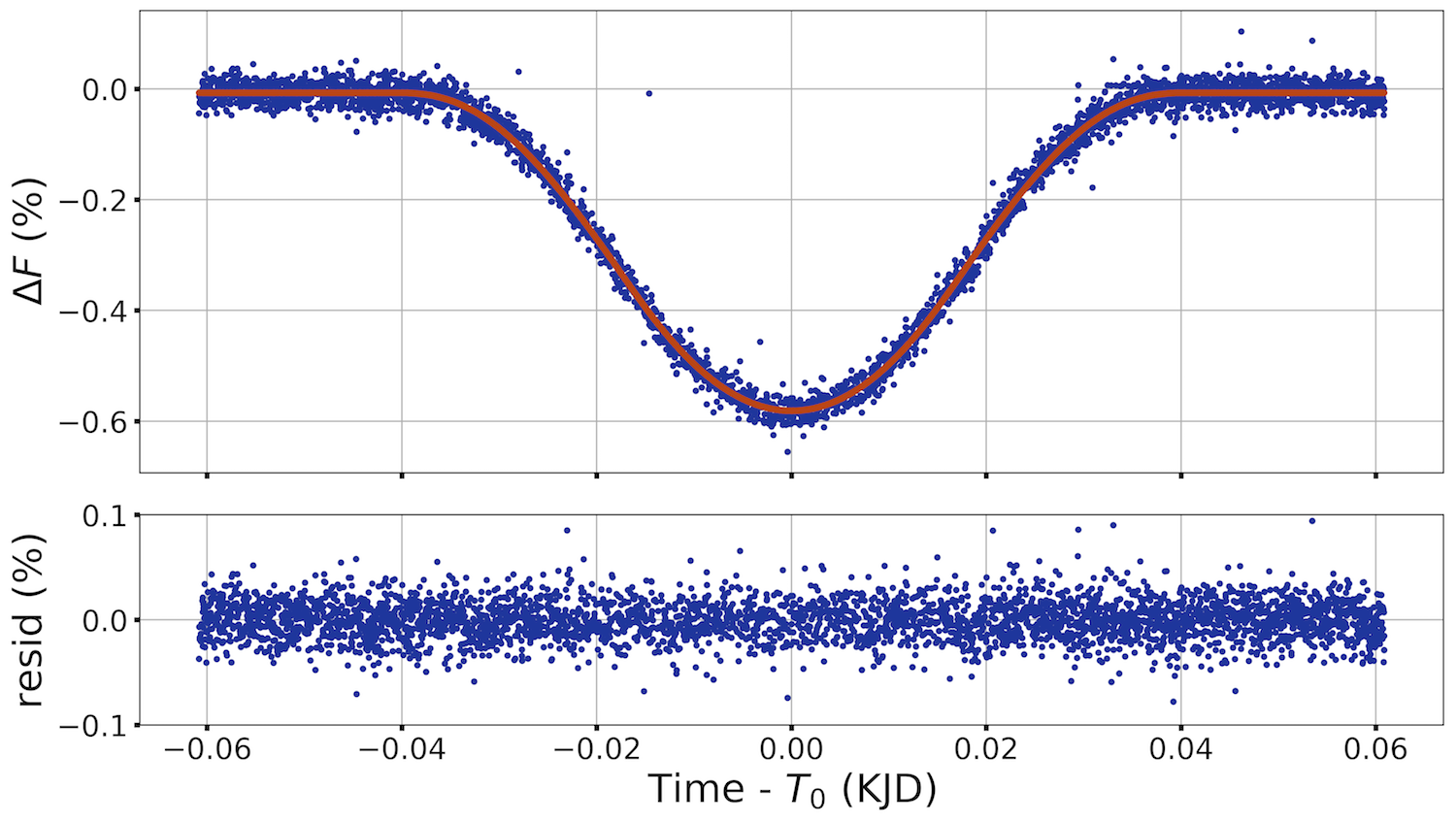}
    \caption{The top panel shows best-fit transit curve for all the in-transit data phase-folded on the ephemeris shown in Figure \ref{fig:TTVs_Analysis_of_Kepler76b}, with residuals between the model (orange line) and data (blue points) in the bottom panel.\label{fig:final_best_fit_transit_Analysis_of_Kepler76b}}
\end{figure}

With this updated ephemeris, we folded together all available data and fit the resulting transit portion of the light curve, with free parameters $a/R_\star$, $b$, $R_{\rm p}/R_\star$, and an estimate of the per-point noise (again using Equation \ref{eqn:likelihood} for the likelihood). We used the same MCMC analysis as above, with 100 walker chains, each with 32,000 links and a burn-in phase of 16,000 links. We aimed for convergence by requiring small auto-correlation times \citep[e.g.,][]{geyer1992} for the mean chain of each fit parameter (none exceeded 250 links) and all Geweke Z-scores \citep{Geweke92evaluatingthe} of about 3 or less. Figure \ref{fig:folded-transit-corner-plot_Analysis-of-Kepler76b} shows the resulting parameter distributions, with the best-fit values and uncertainties given in Table \ref{tbl:Model_Parameters}, and Figure \ref{fig:final_best_fit_transit_Analysis_of_Kepler76b} shows the resulting transit curve and residuals.

\subsection{Searching for Eclipse and Planetary Phase Curve Variability}
\label{sec:Searching}
We turned to a search for orbit-to-orbit variability in the planet's phase curve and secondary eclipse. First, we folded together all 944 orbits-worth of data on our best-fit ephemeris and used the LM algorithm to fit the out-of-transit and eclipse portions (accounting for \kepler's finite integration time) to establish the average fit-values -- the resulting eclipse depth $D = 87 \pm 6$ ppm, a 14.5-$\sigma$ detection. In our search for variability, ideally, we'd analyze the data for each orbit individually to maximize our time resolution. However, with a 14.5-$\sigma$ detection over 944 orbits, we expect that a 3-$\sigma$ detection of the eclipse in the presence of Gaussian noise requires folding together data from about 40 ($=\left( \frac{3}{14.5} \right)^2 \times 944 $) orbits, giving a time resolution of about 60 days. We expect estimates of the parameters $A_{\rm sin}$ and $A_{\rm cos}$ to be more robust, however, since those signals are present throughout most of the orbit (the eclipse only occupies one or two points each orbit). 

For this analysis, we marched a window spanning 10, 20, or 40 orbits from the beginning to the end of the dataset, moving one orbital period at a time. We folded all the out-of-transit points in the window together and fit both Equation \ref{eqn:BEER_curve} and an eclipse, which is represented by the same transit light curve model as above but with a variable depth $D$ and limb-darkening parameters set to zero (i.e., a uniform disk). We used the LM algorithm and took the square root of the diagonal of the covariance matrix scaled by the square root of the resulting reduced $\chi^2$ as the parameter uncertainties -- we compared these uncertainties to those derived from an MCMC analysis for several stacked orbits and found good agreement (to within 1 ppm). 

\begin{figure}
\includegraphics[width=\textwidth]{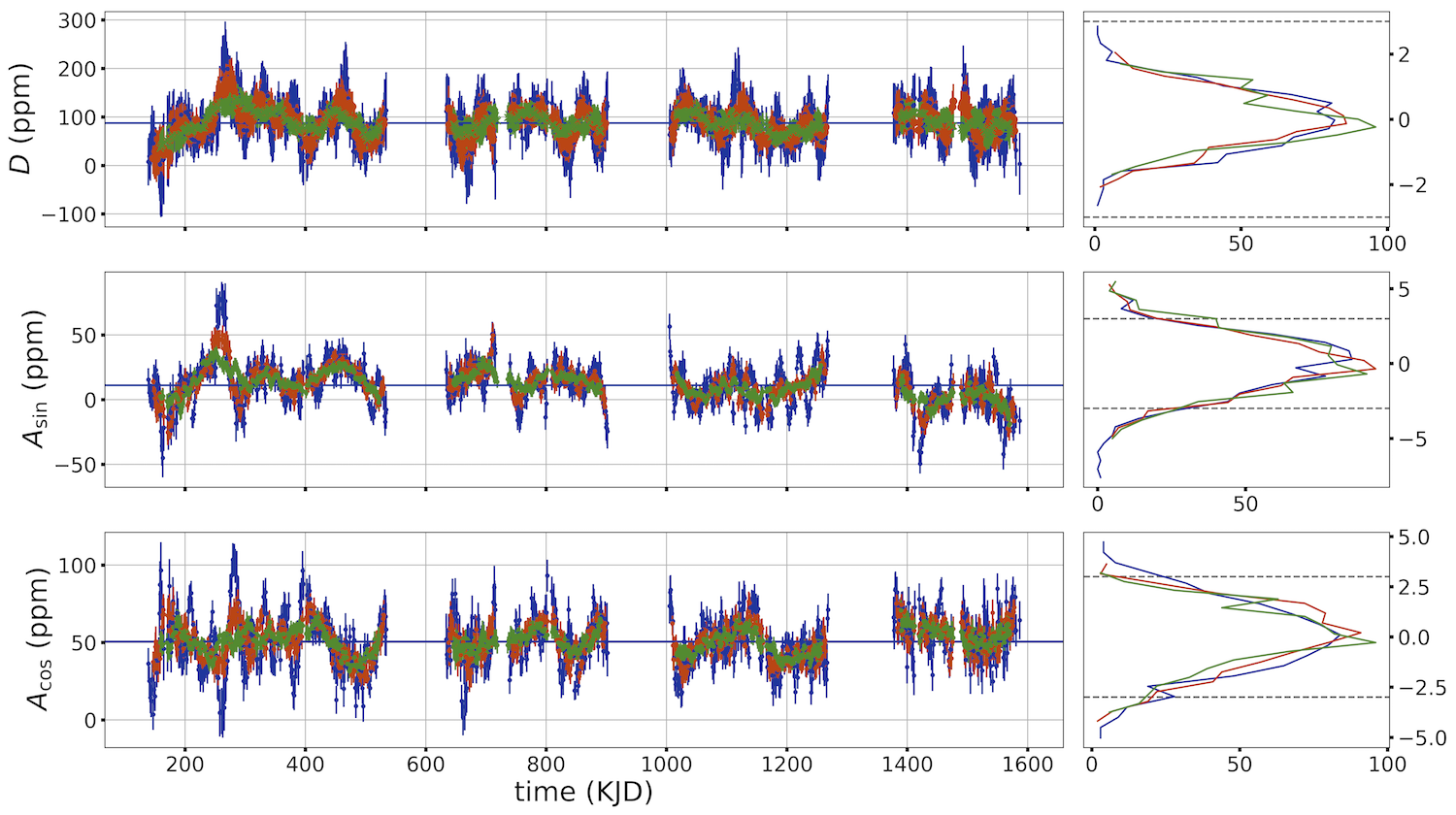}
\caption{The leftmost three panels show variations over time in the eclipse depth $D$ and amplitudes $A_{\rm sin}$ and $A_{\rm cos}$ after stacking and folding data points from consecutive orbits in a window 10 orbits wide (blue points), 20 orbits wide (orange points), and 40 orbits wide (green points). The blue horizontal lines show the best-fit average values for each parameter. The rightmost panels show histograms of how much $D$, $A_{\rm sin}$, and $A_{\rm cos}$ deviate from their respective average values, normalized to their respective uncertainties, and the dashed grey lines illustrate $\pm$3-$\sigma$. \label{fig:planet-phase-curve-var_Analysis_of_Kepler76b}}
\end{figure}

To avoid having our window span large gaps in the data, we required a window to include a large enough number of data points that at least half the desired number of orbits were represented. For example, for the window spanning 10 orbits, we required at least 5 orbital periods worth of data (i.e., about $360 = 5 \times 1.5\ {\rm days}/30\ {\rm min}$ points). We took the mid-eclipse time to be the mid-transit time plus half an orbital period, again assuming zero orbital eccentricity. Figure \ref{fig:planet-phase-curve-var_Analysis_of_Kepler76b} shows the resulting collection of best-fit parameters for the different windows, where the x-coordinate of each point represents the median observational time for all the points in the window. Since the windows span several orbits and the points are only spaced by one orbital period, the parameter fits are not all statistically independent.

The plots seem to show considerable variability, but comparing the variations to the best-fit average values (top right panel of Figure \ref{fig:planet-phase-curve-var_Analysis_of_Kepler76b}) shows that the eclipse depth departs from its average value, 87 ppm, by less than 3-$\sigma$, where $\sigma$ is calculated by adding uncertainties in quadrature \citep[p.~58]{1997ieas.book.....T}. By contrast, $A_{\rm sin}$ and $A_{\rm cos}$ frequently depart by more than 3- and sometimes 5-$\sigma$ from their average values. As expected, though, the more orbits we stacked together, the more the variations are averaged out, but even the points with 40 orbits stacked together (in green) show statistically significant variation. These results, of course, rely on the accuracy of our uncertainty estimates, so we conducted several numerical tests to check them. 

First, we generated and fit several synthetic datasets incorporating Gaussian noise and with the same observational times as the real dataset but holding the phase curve parameters and eclipse depth constant. These fits showed variations visually similar to those in Figure \ref{fig:planet-phase-curve-var_Analysis_of_Kepler76b}, but always the variations in fit parameters were within 3-$\sigma$ of the assumed constant values.

We next gathered and phased up about 7,000 datasets involving 10, 20, or 40 orbits worth of out-of-transit data (10 times the number of points in the left panels in Figure \ref{fig:planet-phase-curve-var_Analysis_of_Kepler76b}) but randomly scattered throughout the whole \kepler\ time-series, i.e., orbits that were not necessarily adjacent in time. We expect that parameter fits to these scattered datasets will show only random scatter, with orbit-to-orbit coherent variations averaged out. We used the Kolmogorov-Smirnov (KS) test \citep[p.~730]{Press:2007:NRE:1403886} to compare the distributions of best-fit parameters (scaled by their respective uncertainties) for these scattered data to the same scaled distributions for the unscattered data (right panels in Figure \ref{fig:planet-phase-curve-var_Analysis_of_Kepler76b}). To judge the range of KS scores we should expect, we randomly sub-sampled the distribution of best-fit parameters for the scattered data 10,000 times and compared these sub-samples to one another. The KS test comparisons indicated that the distribution of eclipse depths for the unscattered data shown in Figure \ref{fig:planet-phase-curve-var_Analysis_of_Kepler76b} resembled closely the distribution for the scattered data, indicating that the variations in the top left panel of Figure \ref{fig:planet-phase-curve-var_Analysis_of_Kepler76b} are consistent with statistical noise. The KS test comparisons for $A_{\rm sin}$ suggests the distribution for the unscattered data is probably inconsistent with random scatter (with a KS probability of about 0.1\%), and the KS test comparisons for $A_{\rm cos}$ are totally inconsistent with random scatter (KS probabilities $<10^{-11}$ when 40 orbits are stacked together). This trend is not surprising. We typically detected the $A_{\rm cos}$ signal at somewhat greater significance than the $A_{\rm sin}$, and both were detected at much greater significance than the eclipse. 

We also employed the same orbit stacking and phasing and fit transit radii (holding other transit parameters fixed) to look for correlations with the phase curve parameters. If the variations in the phase curve arise from some stellar or instrumental effect, we might expect the transit depth to vary in concert; however we found no robust linear correlation.

As a final check, we performed an injection-recovery test for which we generated a synthetic dataset designed to closely mimic the original dataset and included the variations in phase curve parameters shown in Figure \ref{fig:planet-phase-curve-var_Analysis_of_Kepler76b}. For this test, we divided up the full time-series into individual orbits, calculating the mid-transit time for each orbit. We linearly interpolated to that mid-transit time from among the phase curve parameters for 10 orbits stacked together (blue points in Figure \ref{fig:planet-phase-curve-var_Analysis_of_Kepler76b}) to determine the phase curve parameters for the orbit. We generated a phase curve model for that orbit and subtracted it from the original, unconditioned time-series (i.e., from the data shown in the top panel of Figure \ref{fig:raw-conditioned-data_Analysis_of_Kepler76b}). Then, we shifted all the mid-transit times forward in time by about 800 days, i.e., half the maximum observational time in the \kepler\ dataset -- mid-transit times that moved out beyond this maximum time, we wrapped back around to the beginning of the dataset. We next injected synthetic phase curve signals back into these shifted data but this time using the phase curve parameters from the original mid-transit times. This approach has the effect of retaining the time-ordered structure of the phase curve variations while planting the phase curves in a different noise environment. Presumably, if we can recover the phase curve parameters, they are robust against noise. Indeed, after fully re-conditioning these shifted, synthetic data, we were able to recover the vast majority of parameters to within uncertainties. Not surprisingly, for the points not recovered successfully, most were associated with the eclipse depth, which was detected with the least signal-to-noise originally.

\section{Results}
\label{sec:Results}
Our analysis provides an updated set of transit parameters for the Kepler-76 planetary system that was originally announced in \citet{2013ApJ...771...26F}; our ephemeris agrees well with that previous study, while our transit parameters are different. For instance, \citet{2013ApJ...771...26F} estimated $R_{\rm p}/R_\star = 0.0968 \pm 0.0003$, while we estimate $R_{\rm p}/R_\star = 0.085 \pm 0.001$, an 11-$\sigma$ discrepancy. We conducted a variety of tests (ignoring the finite-time \kepler\ integration, applying an alternative data conditioning approach, etc.) and were unable to reproduce the previously reported value. \citet{2019arXiv190101730H} provided equations for calculating the transit depth for a limb-darkened star. Using these equations and the transit parameters given in \citet{2013ApJ...771...26F}, we find that the depth expected (0.66\%) does not seem to agree with the depth of the transit light curve from our study or that shown in Figure 6 of \citet{2013ApJ...771...26F} (about 0.56\%).

In any case, we corroborate detection of planet-induced ellipsoidal variations and the beaming signal, albeit with a different approach and results. Our amplitude for the ellipsoidal variations $A_{\rm ellip}$ is 4-$\sigma$ smaller than that of \citet{2013ApJ...771...26F}. For the beaming signal, we retrieve an amplitude $A_{\rm beam}$ almost 5-$\sigma$ discrepant, although this disagreement may simply arise from our different approach (fitting the planetary phase curve and beaming signals in a non-degenerate way). Using our new transit parameters, we can estimate a new planet-star mass ratio $q$ through the relation $A_{\rm ellip} = \alpha_{\rm ellip} \left( q \sin^2 i \right) \left( R_\star / a \right)^{3}$. In this equation, $\alpha_{\rm ellip}$ depends on the gravity-darkening parameter (0.07 -- \citealp{2011AA...529A..75C}) and linear limb-darkening parameter (0.552 -- \citealp{2015MNRAS.450.1879E}) as described in \citet{1985ApJ...295..143M}. With these values, $\alpha_{\rm ellip} = 1.02$. By re-casting the $A_{\rm ellip}$ equation above in terms of mass ratio $q$ and sampling the distributions of fit parameters from our analysis, we can convert the ellipsoidal signal into $q_{\rm ellip} = \left( 1.9 \pm 0.2 \right) \times 10^{-3}$. In a similar way, we can also estimate a mass ratio for the beaming signal, whose amplitude $A_{\rm beam} = \alpha_{\rm beam} 4 \left( K_{\rm Z} / c \right)$. Taking $\alpha_{\rm beam} = 0.92$, as in \citet{2013ApJ...771...26F}, we find $q_{\rm beam} = \left( 1.7\pm 0.1 \right) \times 10^{-3}$, which is in good agreement with our estimate for $q_{\rm ellip}$. With the stellar mass $M_\star = 1.2 \pm 0.2$ solar masses $M_{\odot}$ from \citet{2013ApJ...771...26F}, these ratios give planetary masses $M_{\rm p, ellip} = 2.4 \pm 0.3$ Jupiter masses ${\rm M_{Jup}}$ and $M_{\rm p, beam} = 2.1 \pm 0.2\ {\rm M_{Jup}}$, respectively. These values agree with the mass estimates in \citet{2013ApJ...771...26F}.

Turning to the results from our phase curve analysis, \emph{a priori}, we expect that the dayside temperature of Kepler-76b, in radiative equilibrium, lies between 2,000 and 2,300 K (between the extremes of complete atmospheric redistribution and a uniform temperature dayside with no redistribution to the nightside), either of which puts a significant fraction of its blackbody curve within the \kepler\ bandpass. If the planet's secondary eclipse represented only thermally emitted light, these temperatures correspond to a depth between 4 and 20 ppm. However, Kepler-76b's average eclipse depth $D = 87 \pm 6$ ppm (within 1.3-$\sigma$ of the result in  \citealt{2013ApJ...771...26F}). With a nightside flux equal to zero, $D$ should be equal to $2\times A_{\rm planet} \cos \left( \phi - \delta \right)$, the planet's phase curve at mid-eclipse. Converting our average $A_{\rm cos}$ and $A_{\rm sin}$-values, we find that these values agree to within 2.3-$\sigma$, consistent with no nightside flux. This eclipse depth corresponds to a brightness temperature of 2,830$^{+50}_{-30}$ K and suggests there is a significant reflected component, $\geq 67$ ppm. Following \citet{2011MNRAS.415.3921F}, we estimate a \kepler\ bandpass-integrated geometric albedo $p_{\rm geo} \geq 0.25$, also consistent with \citet{2013ApJ...771...26F}. 

Finally, we turn to the apparent variations in the planet's phase curve. From Figure \ref{fig:planet-phase-curve-var_Analysis_of_Kepler76b}, variations in $A_{\rm cos}$ and $A_{\rm sin}$ amount to variations of about 40\% in $A_{\rm planet}$ around a median value of $50.5 \pm 1.3$ ppm. Meanwhile, $\delta$ has a median value of $-9^\circ \pm 1.3^\circ$ (an offset east of the sub-stellar point) and swings between about $-40^\circ$ (an eastward offset) and $+20^\circ$ (a westward offset). Figure \ref{fig:Aplanet-delta-var_Analysis_of_Kepler76b} shows the range of $A_{\rm planet}$- and $\delta$-values, along with a linear fit which incorporated the per-point uncertainties in $A_{\rm planet}$ and $\delta$ \citep{boggs1990orthogonal}.

The Kepler-76 system closely resembles the HAT-P-7 system -- both have hot ($>$ 6,000 K) stars hosting large ($>$ 1 ${\rm M_{Jup}}$) short-period planets with observed \kepler\ secondary eclipses indicating dayside brightness temperatures $>$ 2,700 K. As discussed in Section \ref{sec:Introduction}, \citet{2016NatAs...1E...4A} applied a model similar to ours, searching for changes in the planet's phase curve, finding statistically significant fluctuations in $\delta$ between about $6^\circ$ eastward and $9^\circ$ westward of the substellar point over timescales of tens to hundreds of days. That study then applied a model involving atmospheric transport of aerosols and found good qualitative agreement between the model and their observational results. Our results are qualitatively similar, although with much larger apparent swings in $\delta$. In addition, we were able to detect statistically significant variations in the planet's phase curve amplitude, while \citet{2016NatAs...1E...4A} did not.

\section{Discussion and Conclusions}
\label{sec:Discussion}

As discussed in \citet{2016ApJ...828...22P}, the phase curve offset for an eclipsing planet arises, in part, from the competition between thermal emission from the hottest regions and reflection from highly refractive cloud regions on the dayside. Although the daysides of hot Jupiters like Kepler-76b are thought to be too hot for aerosol condensation, super-rotation in their atmospheres can transport nightside aersols across the western terminator onto the dayside before they evaporate. The same super-rotation is thought to shift the hottest region on the dayside eastward of the sub-stellar point. Thus, either with homogeneous clouds or without clouds at all, a hot Jupiter's phase curve would exhibit an eastward (toward $\delta <$ 0) shift. On the other hand, reflective clouds swept over the western terminator would drive the phase curve peak west (toward $\delta > 0$). According to \citet{2016ApJ...828...22P}, the degree of phase shift depends, at least in part, on the cloud composition, with some of the more refractory aerosols able to travel farther east than the less refractory ones. 

\begin{figure}
\includegraphics[width=\textwidth]{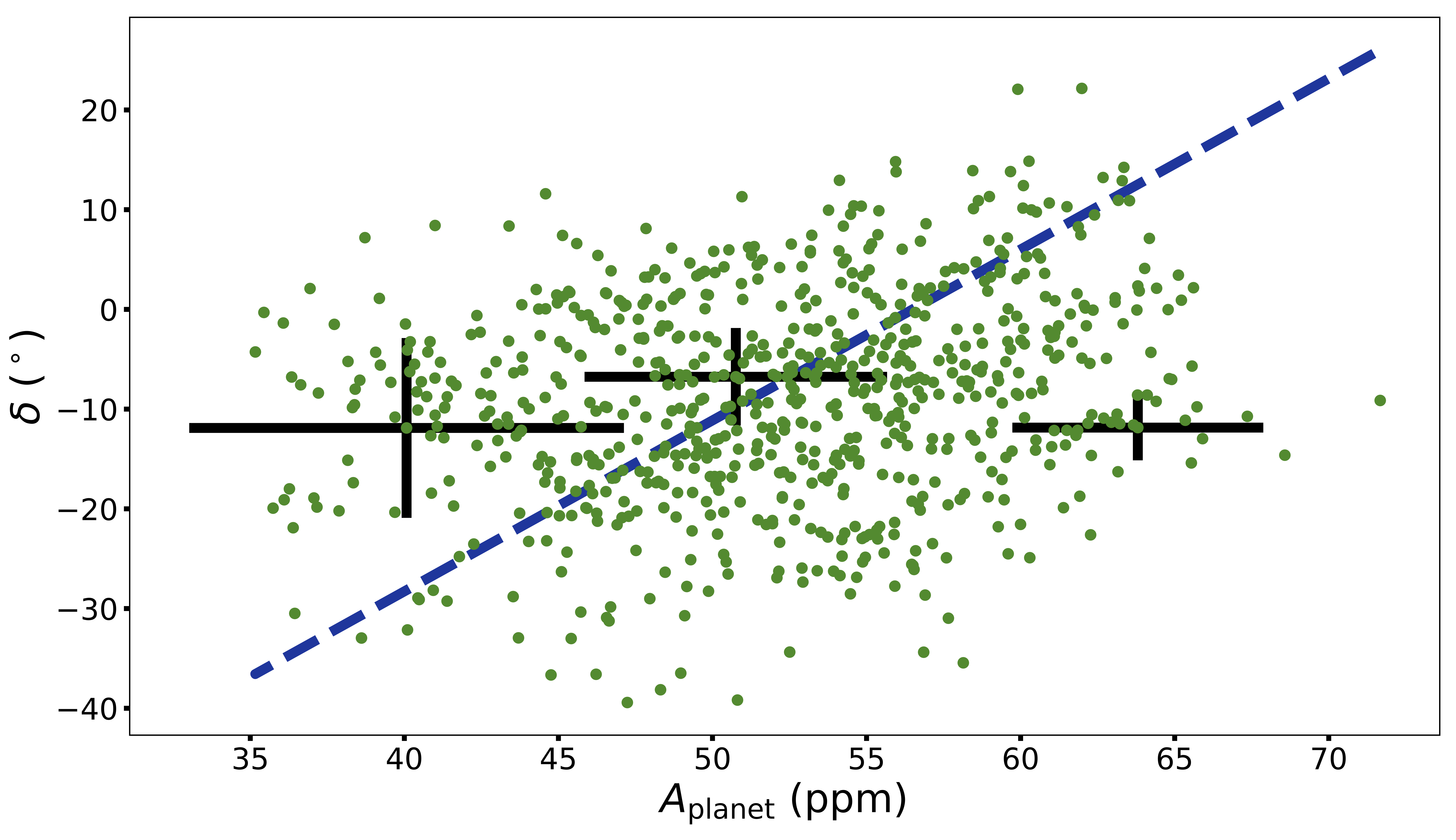}
\caption{The green points show variations in $A_{\rm planet}$ and $\delta$ for 40 orbits stacked together (i.e., based on the green points in Figure \ref{fig:final_best_fit_transit_Analysis_of_Kepler76b}). The black crosses show typical uncertainties, while the dashed blue line shows a linear fit to the green points, with a slope, $m = \left( 1.7 \pm 0.1 \right) ^\circ/{\rm ppm}$. \label{fig:Aplanet-delta-var_Analysis_of_Kepler76b}}
\end{figure}

Variations in the phase curve may help clear up this cloudy story. For planets where reflection from clouds contributes significantly to the phase curve, we might expect a correlation between the phase curve amplitude and offset -- more clouds would produce more signal with a larger westward shift. Figure \ref{fig:Aplanet-delta-var_Analysis_of_Kepler76b} shows a robust (17-$\sigma$) positive correlation between $A_{\rm planet}$ and $\delta$. Since the contributions from clouds and from the dayside hotspot are thought to drive the phase curve peak in opposite directions, their relative influences are encoded in this correlation, and future work with more advanced models for the planetary disk \citep[e.g.,][]{2017ascl.soft11019L} could help to tease them out.

By making some simplifying assumptions about the phase curve, however, we can draw a few tentative conclusions regarding Kepler-76b's atmosphere. As discussed above, a plausible maximum equilibrium temperature expected for Kepler-76b is about 2,300 K, and therefore the maximum contribution to the eclipse is probably about 20 ppm. The other 67 ppm must arise from reflected and/or scattered light. Including Rayleigh scattering and thermal emission, the cloudless atmosphere models from \citet{2016ApJ...828...22P} suggest an effective geometric albedo in the \kepler\ bandpass for Kepler-76b $p_{\rm geo} \approx 0.15$, considerably less than we infer. Thus, some additional component seems to be required. \citet{2016ApJ...828...22P} also indicate that aerosols swept from the nightside and evaporated on the dayside are likely confined to near the western (night-to-day) terminator. If reflection from these aerosols made up the remaining 67 ppm signal for the eclipse, however, we might expect the phase curve to be offset to the west (i.e., $\delta > 0$). This line of reasoning suggests Kepler-76b's dayside is enshrouded by a reflective and homogeneous cloud deck, and the increase in phase curve amplitude and westward shift in offset in Figure \ref{fig:planet-phase-curve-var_Analysis_of_Kepler76b} may arise from periodic arrival of additional aerosols from the nightside.

As discussed in \citet{2016NatAs...1E...4A}, the transport of reflective clouds to the dayside would likely perturb the dayside temperature profiles, potentially reducing the day-night temperature contrast. Since the strength of the super-rotation depends, in part, on this contrast \citep{2011ApJ...738...71S}, a reduction in contrast might reduce both windspeeds and the degree of dayside cloud cover, setting up a feedback. Such a feedback could align with the scenario above, and detailed atmospheric models that include this feedback may help determine the timescales for this feedback for Kepler-76b. In addition to the planet's day-night temperature contrast and rotation rates (which are relatively well constrained), the timescales presumably depend on the optical properties and distribution of the aerosols, which determine their influence on the temperature structure. Thus, comparing the expected to the observed fluctuation timescales could constrain the aerosol properties and mixing in the planet's atmosphere.

\begin{figure}
\includegraphics[width=\textwidth]{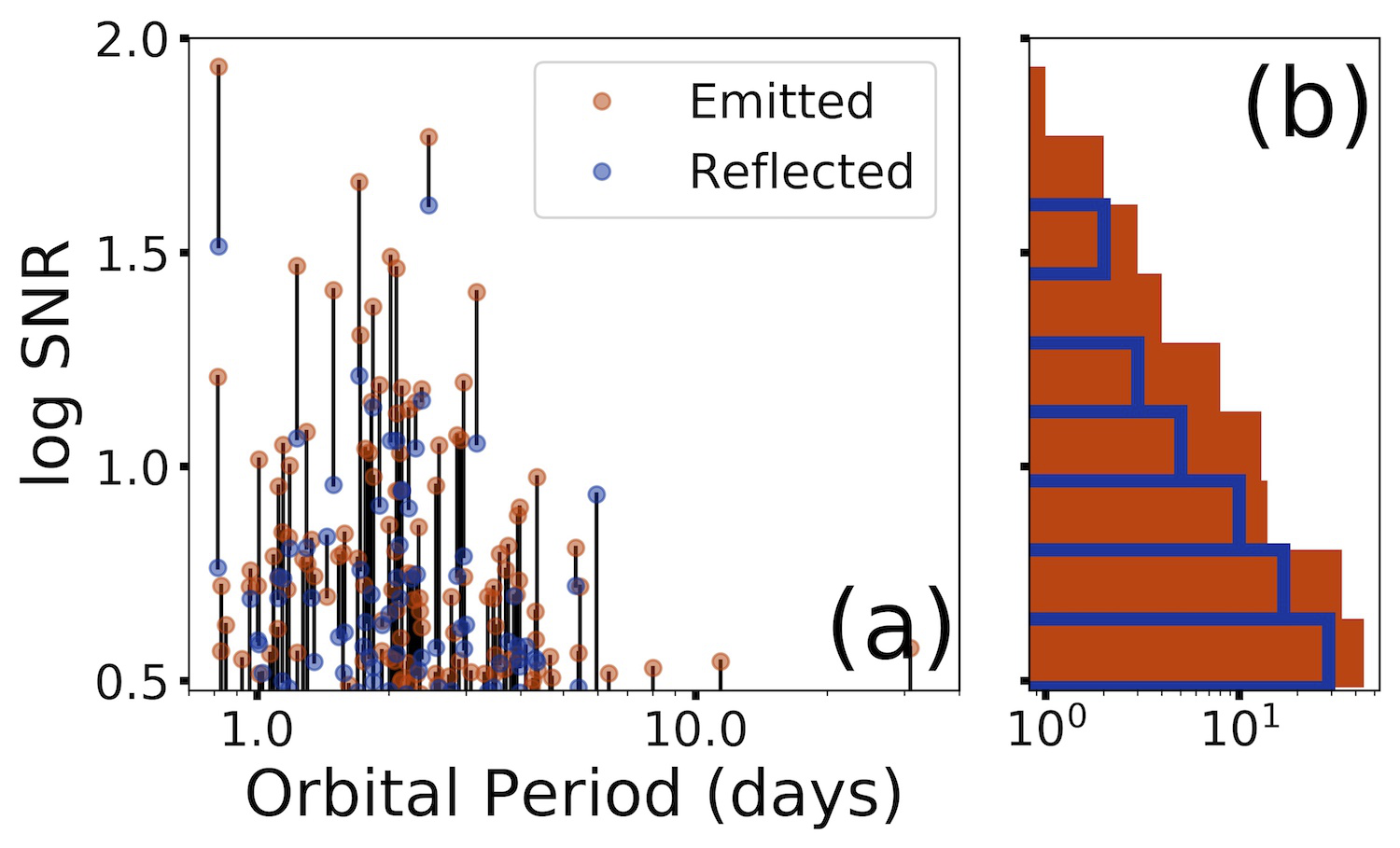}
\caption{The $\log$ of the signal-to-noise ratio $SNR$ for planetary secondary eclipses for a population of synthetic planets from the \tess\ mission yield study of \citet{2018ApJS..239....2B}. The blue dots in (a) show eclipses for perfectly reflecting planets, while the orange dots show eclipses for perfect blackbody planets, with black lines connecting the two eclipses for each planet. Panel (b) shows a histogram of SNR-values. \label{fig:eclipse_estimates}}
\end{figure}

Although the full promise of \kepler\ observations for studying phase curve variations remains unfulfilled, the advent of the \tess\ Mission portends even more insight into planetary atmospheres. Already, \tess\  is detecting phase curves and secondary eclipses from hot Jupiters \citep{2018arXiv181106020S}, and more are expected. For example, Figure \ref{fig:eclipse_estimates} shows estimates of signal-to-noise ratios (SNR) expected for secondary eclipses observed by \tess\ based on the synthetic population of planets in the \tess\ yield calculations from \citet{2018ApJS..239....2B}. For each synthetic planet from that study, we estimated the eclipse depth $D$ for (1) the case in which the planet reflects all of the light it receives from the star (shown in blue) and (2) the case in which the planet emits (and absorbs) light as a perfect blackbody (shown in orange). For the former case, the eclipse depth is given as $\frac{1}{4} \left( \frac{R_{\rm p}}{a} \right)^2$. We then estimated the corresponding SNR by multiplying the transit SNRs given for each planet by planet-star radius ratio squared and then by $D_{\rm reflected}$. For the reflected light eclipse depths, we used the stellar insolation given for each planet and assumed dayside thermal equilibrium with zero albedo for each planet to estimate a brightness temperature and convolved the resulting blackbody curve against the \tess\ spectral response function\footnote{https://tessgi.github.io/TessGiWebsite/the-tess-space-telescope.html}. We combined the insolation and planet's orbital distance to estimate the stellar effective temperature and, likewise, convolved the corresponding blackbody curve against the response function. Finally, we divided the planet's convolved brightness by the star's and used the re-scaled transit SNR to calculate the SNR for the thermally emitted eclipse. Figure \ref{fig:eclipse_estimates} only shows those SNR-values $> 3$, among which the vast majority (54 of 62 total) have $R_{\rm p} \geq 10\ R_{\rm Earth}$. 

The synthetic population from \citet{2018ApJS..239....2B} included 4,553 planets, and Figure \ref{fig:eclipse_estimates} suggests only about 1\% of these may have secondary eclipses detectable by \tess. This result differs from a similar analysis in \citet{2015ApJ...809...77S}, which suggested only 0.01\% of \tess\ planets (about 2) will exhibit observable eclipses, with the difference due to the larger population of short-period, giant planets included in the population we used \citep{2018ApJS..239....2B}. Yet, among those with observable eclipses, our calculation suggests that the eclipse signals will be considerably more sensitive to emitted rather than reflected light, in contrast to the eclipses observed by \kepler; this is largely due to the fact that the \tess\ spectral response function is more sensitive at longer wavelengths than \kepler. Presumably, this different sensitivity means secondary eclipses detectable by \tess\ will reveal atmospheric thermal structures in new ways. However, \tess\ will observe many targets for fewer than 30 days, meaning the kind of variability study presented here would be difficult for all but the most pronounced planetary phase curves or those near the continuous viewing area. The planets with the more easily detectable phase curves, hot Jupiters, also produce the deepest transit signals and are most easily studied from the ground. \citet{2018ApJS..239....2B} estimated that a significant fraction of \tess\ discoveries will be these largest planets, many of which will lie within \tess's continuous viewing zone, providing nearly a year of monitoring. For a planet like Kepler-76b, with an orbital period of 1.5 days, these observations would span hundreds of orbits. Thus, in addition to an exciting menagerie of Earths and super-Earths, \tess\ is poised to reveal a large population of hot Jupiters, potentially ripe for atmospheric characterization. 

\acknowledgments 
This study is based on work supported by NASA under Grant no.~NNX15AB78G issued through the Astrophysical Data Analysis Program by the Science Mission Directorate. The authors gratefully acknowledge assistance from Loretta Cannon in crafting this manuscript and helpful input from Nick Cowan, Lisa Dang, Tamara Rogers, Alexander Hindle, and an anonymous referee.

\facility{Kepler}.
\software{numpy \citep{2011CSE....13b..22V}, astropy \citep{2013A&A...558A..33A, 2018AJ....156..123A}, statsmodels \citep{seabold2010statsmodels}, scipy \citep{scipy}, lightkurve, PyAstronomy, emcee \citep{2013PASP..125..306F}, get\_lds \citep{2015MNRAS.450.1879E}}

\bibliographystyle{aasjournal}
\bibliography{bibliography}

\begin{thebibliography}{}
\expandafter\ifx\csname natexlab\endcsname\relax\def\natexlab#1{#1}\fi
\providecommand{\url}[1]{\href{#1}{#1}}
\providecommand{\dodoi}[1]{doi:~\href{http://doi.org/#1}{\nolinkurl{#1}}}
\providecommand{\doeprint}[1]{\href{http://ascl.net/#1}{\nolinkurl{http://ascl.net/#1}}}
\providecommand{\doarXiv}[1]{\href{https://arxiv.org/abs/#1}{\nolinkurl{https://arxiv.org/abs/#1}}}

\bibitem[{{Armstrong} {et~al.}(2016){Armstrong}, {de Mooij}, {Barstow},
  {Osborn}, {Blake}, \& {Saniee}}]{2016NatAs...1E...4A}
{Armstrong}, D.~J., {de Mooij}, E., {Barstow}, J., {et~al.} 2016, Nature
  Astronomy, 1, 0004, \dodoi{10.1038/s41550-016-0004}

\bibitem[{{Astropy Collaboration} {et~al.}(2013){Astropy Collaboration},
  {Robitaille}, {Tollerud}, {Greenfield}, {Droettboom}, {Bray}, {Aldcroft},
  {Davis}, {Ginsburg}, {Price-Whelan}, {Kerzendorf}, {Conley}, {Crighton},
  {Barbary}, {Muna}, {Ferguson}, {Grollier}, {Parikh}, {Nair}, {Unther},
  {Deil}, {Woillez}, {Conseil}, {Kramer}, {Turner}, {Singer}, {Fox}, {Weaver},
  {Zabalza}, {Edwards}, {Azalee Bostroem}, {Burke}, {Casey}, {Crawford},
  {Dencheva}, {Ely}, {Jenness}, {Labrie}, {Lim}, {Pierfederici}, {Pontzen},
  {Ptak}, {Refsdal}, {Servillat}, \& {Streicher}}]{2013A&A...558A..33A}
{Astropy Collaboration}, {Robitaille}, T.~P., {Tollerud}, E.~J., {et~al.} 2013,
  \aap, 558, A33, \dodoi{10.1051/0004-6361/201322068}

\bibitem[{{Astropy Collaboration} {et~al.}(2018){Astropy Collaboration},
  {Price-Whelan}, {Sip{\H o}cz}, {G{\"u}nther}, {Lim}, {Crawford}, {Conseil},
  {Shupe}, {Craig}, {Dencheva}, {Ginsburg}, {VanderPlas}, {Bradley},
  {P{\'e}rez-Su{\'a}rez}, {de Val-Borro}, {Aldcroft}, {Cruz}, {Robitaille},
  {Tollerud}, {Ardelean}, {Babej}, {Bach}, {Bachetti}, {Bakanov}, {Bamford},
  {Barentsen}, {Barmby}, {Baumbach}, {Berry}, {Biscani}, {Boquien}, {Bostroem},
  {Bouma}, {Brammer}, {Bray}, {Breytenbach}, {Buddelmeijer}, {Burke},
  {Calderone}, {Cano Rodr{\'{\i}}guez}, {Cara}, {Cardoso}, {Cheedella},
  {Copin}, {Corrales}, {Crichton}, {D'Avella}, {Deil}, {Depagne}, {Dietrich},
  {Donath}, {Droettboom}, {Earl}, {Erben}, {Fabbro}, {Ferreira}, {Finethy},
  {Fox}, {Garrison}, {Gibbons}, {Goldstein}, {Gommers}, {Greco}, {Greenfield},
  {Groener}, {Grollier}, {Hagen}, {Hirst}, {Homeier}, {Horton}, {Hosseinzadeh},
  {Hu}, {Hunkeler}, {Ivezi{\'c}}, {Jain}, {Jenness}, {Kanarek}, {Kendrew},
  {Kern}, {Kerzendorf}, {Khvalko}, {King}, {Kirkby}, {Kulkarni}, {Kumar},
  {Lee}, {Lenz}, {Littlefair}, {Ma}, {Macleod}, {Mastropietro}, {McCully},
  {Montagnac}, {Morris}, {Mueller}, {Mumford}, {Muna}, {Murphy}, {Nelson},
  {Nguyen}, {Ninan}, {N{\"o}the}, {Ogaz}, {Oh}, {Parejko}, {Parley}, {Pascual},
  {Patil}, {Patil}, {Plunkett}, {Prochaska}, {Rastogi}, {Reddy Janga},
  {Sabater}, {Sakurikar}, {Seifert}, {Sherbert}, {Sherwood-Taylor}, {Shih},
  {Sick}, {Silbiger}, {Singanamalla}, {Singer}, {Sladen}, {Sooley},
  {Sornarajah}, {Streicher}, {Teuben}, {Thomas}, {Tremblay}, {Turner},
  {Terr{\'o}n}, {van Kerkwijk}, {de la Vega}, {Watkins}, {Weaver}, {Whitmore},
  {Woillez}, {Zabalza}, \& {Astropy Contributors}}]{2018AJ....156..123A}
{Astropy Collaboration}, {Price-Whelan}, A.~M., {Sip{\H o}cz}, B.~M., {et~al.}
  2018, \aj, 156, 123, \dodoi{10.3847/1538-3881/aabc4f}

\bibitem[{{Barclay} {et~al.}(2018){Barclay}, {Pepper}, \&
  {Quintana}}]{2018ApJS..239....2B}
{Barclay}, T., {Pepper}, J., \& {Quintana}, E.~V. 2018, \apjs, 239, 2,
  \dodoi{10.3847/1538-4365/aae3e9}

\bibitem[{Boggs \& Rogers(1990)}]{boggs1990orthogonal}
Boggs, P.~T., \& Rogers, J.~E. 1990, Contemporary Mathematics, 112, 183

\bibitem[{{Charbonneau} {et~al.}(2005){Charbonneau}, {Allen}, {Megeath},
  {Torres}, {Alonso}, {Brown}, {Gilliland}, {Latham}, {Mandushev}, {O'Donovan},
  \& {Sozzetti}}]{2005ApJ...626..523C}
{Charbonneau}, D., {Allen}, L.~E., {Megeath}, S.~T., {et~al.} 2005, \apj, 626,
  523, \dodoi{10.1086/429991}

\bibitem[{{Claret} \& {Bloemen}(2011)}]{2011AA...529A..75C}
{Claret}, A., \& {Bloemen}, S. 2011, \aap, 529, A75,
  \dodoi{10.1051/0004-6361/201116451}

\bibitem[{{Dang} {et~al.}(2018){Dang}, {Cowan}, {Schwartz}, {Rauscher},
  {Zhang}, {Knutson}, {Line}, {Dobbs-Dixon}, {Deming}, {Sundararajan},
  {Fortney}, \& {Zhao}}]{2018NatAs...2..220D}
{Dang}, L., {Cowan}, N.~B., {Schwartz}, J.~C., {et~al.} 2018, Nature Astronomy,
  2, 220, \dodoi{10.1038/s41550-017-0351-6}

\bibitem[{{Deming} {et~al.}(2005){Deming}, {Seager}, {Richardson}, \&
  {Harrington}}]{2005Natur.434..740D}
{Deming}, D., {Seager}, S., {Richardson}, L.~J., \& {Harrington}, J. 2005,
  \nat, 434, 740, \dodoi{10.1038/nature03507}

\bibitem[{{Demory} {et~al.}(2016){Demory}, {Gillon}, {de Wit}, {Madhusudhan},
  {Bolmont}, {Heng}, {Kataria}, {Lewis}, {Hu}, {Krick}, {Stamenkovi{\'c}},
  {Benneke}, {Kane}, \& {Queloz}}]{2016Natur.532..207D}
{Demory}, B.-O., {Gillon}, M., {de Wit}, J., {et~al.} 2016, \nat, 532, 207,
  \dodoi{10.1038/nature17169}

\bibitem[{{Espinoza} \& {Jord{\'a}n}(2015)}]{2015MNRAS.450.1879E}
{Espinoza}, N., \& {Jord{\'a}n}, A. 2015, \mnras, 450, 1879,
  \dodoi{10.1093/mnras/stv744}

\bibitem[{{Esteves} {et~al.}(2015){Esteves}, {De Mooij}, \&
  {Jayawardhana}}]{2015ApJ...804..150E}
{Esteves}, L.~J., {De Mooij}, E.~J.~W., \& {Jayawardhana}, R. 2015, \apj, 804,
  150, \dodoi{10.1088/0004-637X/804/2/150}

\bibitem[{{Faigler} \& {Mazeh}(2011)}]{2011MNRAS.415.3921F}
{Faigler}, S., \& {Mazeh}, T. 2011, \mnras, 415, 3921,
  \dodoi{10.1111/j.1365-2966.2011.19011.x}

\bibitem[{{Faigler} {et~al.}(2013){Faigler}, {Tal-Or}, {Mazeh}, {Latham}, \&
  {Buchhave}}]{2013ApJ...771...26F}
{Faigler}, S., {Tal-Or}, L., {Mazeh}, T., {Latham}, D.~W., \& {Buchhave}, L.~A.
  2013, \apj, 771, 26, \dodoi{10.1088/0004-637X/771/1/26}

\bibitem[{{Foreman-Mackey} {et~al.}(2013){Foreman-Mackey}, {Hogg}, {Lang}, \&
  {Goodman}}]{2013PASP..125..306F}
{Foreman-Mackey}, D., {Hogg}, D.~W., {Lang}, D., \& {Goodman}, J. 2013, \pasp,
  125, 306, \dodoi{10.1086/670067}

\bibitem[{{Geweke}(1992)}]{Geweke92evaluatingthe}
{Geweke}, J. 1992, in IN BAYESIAN STATISTICS (University Press), 169--193

\bibitem[{Geyer(1992)}]{geyer1992}
Geyer, C.~J. 1992, Statist. Sci., 7, 473, \dodoi{10.1214/ss/1177011137}

\bibitem[{{Heller}(2019)}]{2019arXiv190101730H}
{Heller}, R. 2019, arXiv e-prints.
\newblock \doarXiv{1901.01730}

\bibitem[{{Hindle} {et~al.}(2019){Hindle}, {Bushby}, \&
  {Rogers}}]{2019ApJ...872L..27H}
{Hindle}, A.~W., {Bushby}, P.~J., \& {Rogers}, T.~M. 2019, \apjl, 872, L27,
  \dodoi{10.3847/2041-8213/ab05dd}

\bibitem[{Jones {et~al.}(2001--)Jones, Oliphant, Peterson, {et~al.}}]{scipy}
Jones, E., Oliphant, T., Peterson, P., {et~al.} 2001--, {SciPy}: Open source
  scientific tools for {Python}.
\newblock \url{http://www.scipy.org/}

\bibitem[{{Kipping}(2010)}]{2010MNRAS.408.1758K}
{Kipping}, D.~M. 2010, \mnras, 408, 1758,
  \dodoi{10.1111/j.1365-2966.2010.17242.x}

\bibitem[{{Knutson} {et~al.}(2007){Knutson}, {Charbonneau}, {Allen}, {Fortney},
  {Agol}, {Cowan}, {Showman}, {Cooper}, \& {Megeath}}]{2007Natur.447..183K}
{Knutson}, H.~A., {Charbonneau}, D., {Allen}, L.~E., {et~al.} 2007, \nat, 447,
  183, \dodoi{10.1038/nature05782}

\bibitem[{{Koll} \& {Komacek}(2018)}]{2018ApJ...853..133K}
{Koll}, D.~D.~B., \& {Komacek}, T.~D. 2018, \apj, 853, 133,
  \dodoi{10.3847/1538-4357/aaa3de}

\bibitem[{{Loeb} \& {Gaudi}(2003)}]{2003ApJ...588L.117L}
{Loeb}, A., \& {Gaudi}, B.~S. 2003, \apjl, 588, L117, \dodoi{10.1086/375551}

\bibitem[{{Lomb}(1976)}]{1976Ap&SS..39..447L}
{Lomb}, N.~R. 1976, \apss, 39, 447, \dodoi{10.1007/BF00648343}

\bibitem[{{Louden} \& {Kreidberg}(2017)}]{2017ascl.soft11019L}
{Louden}, T., \& {Kreidberg}, L. 2017, {SPIDERMAN: Fast code to simulate
  secondary transits and phase curves}, Astrophysics Source Code Library.
\newblock \doeprint{1711.019}

\bibitem[{{Mandel} \& {Agol}(2002)}]{2002ApJ...580L.171M}
{Mandel}, K., \& {Agol}, E. 2002, \apjl, 580, L171, \dodoi{10.1086/345520}

\bibitem[{{Morris}(1985)}]{1985ApJ...295..143M}
{Morris}, S.~L. 1985, \apj, 295, 143, \dodoi{10.1086/163359}

\bibitem[{Newville {et~al.}(2014)Newville, Stensitzki, Allen, \&
  Ingargiola}]{newville_2014_11813}
Newville, M., Stensitzki, T., Allen, D.~B., \& Ingargiola, A. 2014, {LMFIT:
  Non-Linear Least-Square Minimization and Curve-Fitting for Python},
  \dodoi{10.5281/zenodo.11813}.
\newblock \url{https://doi.org/10.5281/zenodo.11813}

\bibitem[{{Parmentier} {et~al.}(2016){Parmentier}, {Fortney}, {Showman},
  {Morley}, \& {Marley}}]{2016ApJ...828...22P}
{Parmentier}, V., {Fortney}, J.~J., {Showman}, A.~P., {Morley}, C., \&
  {Marley}, M.~S. 2016, \apj, 828, 22, \dodoi{10.3847/0004-637X/828/1/22}

\bibitem[{{Parmentier} {et~al.}(2013){Parmentier}, {Showman}, \&
  {Lian}}]{2013A&A...558A..91P}
{Parmentier}, V., {Showman}, A.~P., \& {Lian}, Y. 2013, \aap, 558, A91,
  \dodoi{10.1051/0004-6361/201321132}

\bibitem[{Press {et~al.}(2007)Press, Teukolsky, Vetterling, \&
  Flannery}]{Press:2007:NRE:1403886}
Press, W.~H., Teukolsky, S.~A., Vetterling, W.~T., \& Flannery, B.~P. 2007,
  Numerical Recipes 3rd Edition: The Art of Scientific Computing, 3rd edn. (New
  York, NY, USA: Cambridge University Press)

\bibitem[{{Scargle}(1982)}]{1982ApJ...263..835S}
{Scargle}, J.~D. 1982, \apj, 263, 835, \dodoi{10.1086/160554}

\bibitem[{Seabold \& Perktold(2010)}]{seabold2010statsmodels}
Seabold, S., \& Perktold, J. 2010, in 9th Python in Science Conference

\bibitem[{{Showman} \& {Guillot}(2002)}]{2002A&A...385..166S}
{Showman}, A.~P., \& {Guillot}, T. 2002, \aap, 385, 166,
  \dodoi{10.1051/0004-6361:20020101}

\bibitem[{{Showman} \& {Polvani}(2011)}]{2011ApJ...738...71S}
{Showman}, A.~P., \& {Polvani}, L.~M. 2011, \apj, 738, 71,
  \dodoi{10.1088/0004-637X/738/1/71}

\bibitem[{{Shporer} {et~al.}(2018){Shporer}, {Wong}, {Huang}, {Line},
  {Stassun}, {Fetherolf}, {Kane}, {Ricker}, {Latham}, {Seager}, {Winn},
  {Jenkins}, {Glidden}, {Berta-Thompson}, {Ting}, {Li}, \&
  {Haworth}}]{2018arXiv181106020S}
{Shporer}, A., {Wong}, I., {Huang}, C.~X., {et~al.} 2018, arXiv e-prints.
\newblock \doarXiv{1811.06020}

\bibitem[{{Sing} {et~al.}(2016){Sing}, {Fortney}, {Nikolov}, {Wakeford},
  {Kataria}, {Evans}, {Aigrain}, {Ballester}, {Burrows}, {Deming},
  {D{\'e}sert}, {Gibson}, {Henry}, {Huitson}, {Knutson}, {Lecavelier Des
  Etangs}, {Pont}, {Showman}, {Vidal-Madjar}, {Williamson}, \&
  {Wilson}}]{2016Natur.529...59S}
{Sing}, D.~K., {Fortney}, J.~J., {Nikolov}, N., {et~al.} 2016, \nat, 529, 59,
  \dodoi{10.1038/nature16068}

\bibitem[{{Sudarsky} {et~al.}(2000){Sudarsky}, {Burrows}, \&
  {Pinto}}]{2000ApJ...538..885S}
{Sudarsky}, D., {Burrows}, A., \& {Pinto}, P. 2000, \apj, 538, 885,
  \dodoi{10.1086/309160}

\bibitem[{{Sullivan} {et~al.}(2015){Sullivan}, {Winn}, {Berta-Thompson},
  {Charbonneau}, {Deming}, {Dressing}, {Latham}, {Levine}, {McCullough},
  {Morton}, {Ricker}, {Vanderspek}, \& {Woods}}]{2015ApJ...809...77S}
{Sullivan}, P.~W., {Winn}, J.~N., {Berta-Thompson}, Z.~K., {et~al.} 2015, \apj,
  809, 77, \dodoi{10.1088/0004-637X/809/1/77}

\bibitem[{{Taylor}(1997)}]{1997ieas.book.....T}
{Taylor}, J. 1997, {Introduction to Error Analysis, the Study of Uncertainties
  in Physical Measurements, 2nd Edition} (University Science Books)

\bibitem[{{van der Walt} {et~al.}(2011){van der Walt}, {Colbert}, \&
  {Varoquaux}}]{2011CSE....13b..22V}
{van der Walt}, S., {Colbert}, S.~C., \& {Varoquaux}, G. 2011, Computing in
  Science and Engineering, 13, 22, \dodoi{10.1109/MCSE.2011.37}

\bibitem[{{Welsh} {et~al.}(2010){Welsh}, {Orosz}, {Seager}, {Fortney},
  {Jenkins}, {Rowe}, {Koch}, \& {Borucki}}]{2010ApJ...713L.145W}
{Welsh}, W.~F., {Orosz}, J.~A., {Seager}, S., {et~al.} 2010, \apjl, 713, L145,
  \dodoi{10.1088/2041-8205/713/2/L145}

\end{thebibliography}

\end{document}